# A SYSTEMATIC STUDY OF VARIOUS FINGERTIP DETECTION TECHNIQUES FOR AIR WRITING USING MACHINE LEARNING

Heena[1]

*PhD Scholar, CT University, Jagraon, Punjab, India*

[1]*heenajand21@gmail.com*

Dr. Sandeep Ranjan[2]

*Professor, CT University, Jagraon, Punjab, India*

[2]*ersandeepranjan@yahoo.com*

*Abstract* —The recent advancement in technology breaks the barriers to communication between users and computers. The communication between humans and computers includes emotion and gesture recognition. Emotions can be recognized on the face of humans whereas gesture recognition includes hand and body gesture recognition. Fingertip detection is also part of it. Gesture recognition is the way of interaction that is used in air writing. Users can control the devices with simple gestures without touching them. It is how computers can understand human language which will reduce the interaction barriers between them. This paper discusses the different techniques that can be used for fingertip detection in air writing using machine learning.

## I. INTRODUCTION

One of the fundamental technologies for gesture recognition is fingertip detection, and it significantly affects how well gestures are identified. The oldest point of a-finger detection requires the employment of supplemental technology, such as data gloves to get information about hand and fingertip recognition, but these gadgets are pricey and frequently have a limited detecting range. As a result of the growth of technology has produced breakthroughs in image processing precise fingertip detection using fundamental image analysis using processing methods like histogram segmentation, skin color segmentation technique, and technology. However, each of these techniques depends on the handling of color information to get knowledge about fingertips, or the outcome depends on the lighting circumstances of fingertip detection, and it can't work in the situation [2].

Hand gesture recognition is a technique for the mathematical interpretation of hand movements by computers. It can be used to facilitate the interaction between humans and computers [16]. Gesture recognition is effective for applications such as device control, entertainment, health care, and education. Hand gesture recognition approaches can be separated into two classes: Vision-Based Recognition (VBR) algorithms and Sensor-Based Recognition (SBR) algorithms. The VBR algorithms perform gesture recognition from images captured by a camera. Although accurate classification is possible, high computation efforts may be required to extract information from images for both training and inference operations [26]. The SBR algorithms are based on sensors other than a camera. Commonly used sensors include accelerometers, gyroscopes, photoplethysmography (PPG), flex sensors, electromyography (EMG), Radio Frequency Identification (RFID), Wi-Fi Channel State Information (CSI), and the fusion of these sensors. In some SBR-based studies, high classification accuracy for gesture recognition has been observed. However, many of these techniques do not support the recognition of a sequence of gestures. Only isolated gestures can be recognized [19].

Furthermore, some SBR techniques are based on Dynamic Time Warping (DTW) technique for gesture recognition, which may incur high computation complexities. The Recurrent Neural Networks (RNNs) and their variants such as Long Short-Term Memory (LSTM) algorithm are effective for the recognition of gesture sequences with low computation costs. Nevertheless, because of the inherent gradient vanishing problem, the algorithms may not be able to effectively exploit the long-term dependency of the sensory data. The dependency may be beneficial for accurate recognition when slow gesture movements are observed. In addition to RNNs, feedforward networks such as Convolutional Neural Networks (CNNs) can be used for hand gesture recognition. The One-Dimensional (1D) CNNs are effective for several applications such as the classifications of ECG signals, human activities, and internet traffic. Gesture recognition based on basic 1D CNNs may achieve high classification accuracy when the kernel sizes and/or the depth of the network are large. However, in these cases, the computation complexities for inference may also be high. Although the computation load can be alleviated by lowering the kernel sizes and/or the depth of the network, the coverage area of the receptive field will then become small. Consequently, the classification error may become large because the long-term dependency may not be effectively exploited with a small receptive field. The WaveNet can be used to solve the problem. It is based on dilated convolution operations for the growth of receptive fields with low computational complexities. However, WaveNet is used for signal generation with additional autoregression operations. Direct applications of WaveNet to hand gesture recognition may then be difficult.

A lot of techniques are available for fingertip detection. A few of them have been studied and explained in this paper.

## II. VARIOUS FINGERTIP DETECTION TECHNIQUES
### A. Using Web Camera

The use of fingertips to track or control motions in many vision-based applications is common. Given that humans convey their emotions most frequently through hand gestures, gesture identification is a natural approach to transmitting messages to the machine [12]. To identify the fingertip using a camera in real-time video then single or multiple cameras can be used. In this, Video is a fixed-rate succession of visual frames.

B. *Using LED or color marker for finger movement*

It is a technique in which an LED is mounted on the user's finger and the web camera is used to track the finger. The character drawn is compared with that present in the database. It returns the alphabet that matches the pattern drawn [12]. It requires a red-colored LED pointed light source attached to the finger. Also, it is assumed that there is no red-colored object other than the LED light within the focus of the web camera.

C. *Accelerometer and gyroscope sensors*

Accelerometer is an automatic tool for measuring acceleration, detecting and measuring vibration, and measuring acceleration due to the body's inclination. The accelerometer can be used to measure vibrations in automobile vehicles, engines, buildings, and security installations. An accelerometer can also be used in measuring acceleration in electronic equipment, like 3-dimensional games, computer mouse and telephones and earthquake activity, etc. Acceleration makes a state on the velocity with time [1]. If the speed on range time is increased then it is called acceleration. But if the speed decreased from the previous speed, then it's called deceleration. The change in the movement of the object's direction will also cause acceleration. Newton's Second law of motion says that the acceleration (m/s2) of a body is directly proportional to the net force (Newton) acting on the body, and inversely proportional to its mass (gram) [17].

Gyroscope is a device that can sense the velocity of angular motion of the frame where it is mounted if that frame is rotated. The classes of gyroscopes depend on their physical and technological operations. The gyroscope can be used individually or used for something complex systems, like Inertial Measurement Unit (IMU), gyrocompass, attitude heading reference system, and navigation system [4].

D. *Leap Motion Detector*

The Leap Motion controller is a cutting-edge hand gesture recognition tool with sub-millimeter accuracy. The Leap Motion sensor can be thought of as a stereo vision-based optical tracking device. Its Software Development Kit (SDK) provides data about the rotations of the hand (such as Roll, Pitch, and Yaw), as well as data about the fingertips, pen tip, hand palm position, and other predefined objects in the Cartesian space. Every delivered position is concerning the center of the Leap Motion Controller [8][14].

E. *Kinect Sensors*

Kinect is a Microsoft-developed somatosensory external device for the XBOX360 that was released in 2010. The three eyes are Kinect's most significant practical use, followed by the infrared camera, the color camera, and the infrared depth camera. The infrared camera and infrared depth camera produce the depth image, and fingertip detection based on the depth image can cut through interference from light and other objects [3]. After rendering the depth information, the information about human body joints is collected, and by using the joints of the hand to monitor the position of the hand, the fingertips are detected in complex backgrounds and movements [13][15].

### F. Raspberry Pi Board

A single-board computer the size of a credit card is called the Raspberry Pi. The Raspberry Pi Foundation is responsible for its development in the UK. The target audience is primarily students. Through licensed manufacturing agreements with the businesses Newark element14 (Premier Farnell), RS Components, and Egoman, the Raspberry Pi is produced in two board configurations. The Raspberry Pi is offered online by these businesses [6]. The red color and absence of FCC/CE marks set apart the version made by Egoman for distribution only in China and Taiwan. Every manufacturer uses the same hardware. The Broadcom (BCM2836) system on a chip (SoC) that powers the Raspberry Pi2 has an ARM (900 MHz) CPU. It has no internal hard drive or solid-state drive, so it utilizes an SD card for persistent storage and booting. Numerous Operating Systems are compatible with Pi boards. Many programming languages are supported by it, including Python, C, C++, Java, Scratch, and Ruby, which are all preinstalled on the Raspberry Pi [7].

### G. Micaz Mote

It is a low-cost application and can be used by disabled persons who do not have limbs or fingers. Patients just have to wear the mote as a wristwatch or like a bracelet and move their hands for writing characters. This solution has in-built sensors like an accelerometer and gyroscope. It mainly detects 8 sets of strokes: ← ↑ ↓ ↘ ↙ ↗ ↖ → These are called strokes (vertical and horizontal strokes). When the user moves the hand, the accelerometer starts increasing, and when the hand stops, the accelerometer decreases and come to zero. The photosensor which is the part of the sensor board is used to take the indication from the user. When the user finishes the character, he put his hand on the photosensor in this way it blocks the light on the photosensor and its value goes down. The system understands that here is the end of the character and prints it on the screen. ← Stroke can also be used for erasing the character in case of a typing error. This system can detect English characters and numbers as well [5].

### III. CONCLUSION

Building effective human-machine interactions depends on gesture recognition. In this paper, we have studied the different techniques for fingertip detection that can be used in air writing. For writing characters in the air all of these techniques are useful but all have their drawbacks also. The results of these techniques depend on various factors like the writing speed of a particular subject, writing style, etc that must be taken care of while working on it. There can be lots of other techniques like YOLO, scan fitting algorithm, and LSTM that can also be useful in fingertip detection. While working on Fingertip detection, one can select any of above mention techniques or algorithms or combine them to find the best results. We can utilize tracking techniques like Kalman filters and Extended Kalman filters to enhance hand detection technology.